\documentclass[3p,times,procedia]{elsarticle}

\usepackage{ecrc}

\volume{00}

\firstpage{1}

\journalname{Physics Procedia}

\runauth{}

\jid{phpro}

\jnltitlelogo{Physics Procedia}

\CopyrightLine{2011}{Published by Elsevier Ltd.}

\usepackage{amssymb}

\usepackage[figuresright]{rotating}

\usepackage[english]{babel}
\usepackage{amsmath, amssymb}
\usepackage{graphicx}
\usepackage{subfigure}
\usepackage{color}
\usepackage{xspace}
\usepackage{multirow}
\usepackage{xspace}

\newcommand{\pythia}{{\ttfamily PYTHIA}\xspace}
\newcommand{\herwig}{{\ttfamily HERWIG}\xspace}
\newcommand{\herwigpp}{{\ttfamily HERWIG++}\xspace}

\begin{document}

\begin{frontmatter}



\dochead{}

\title{Gamma-ray and neutrino fluxes from Heavy Dark Matter\\ in the Galactic Center}


\author[a]{V.~Gammaldi\footnote{gammaldi@pas.ucm.es}}
\author[a]{J.~A.~R.~Cembranos}
\author[a]{A.~de~la~Cruz-Dombriz}
\author[b]{R.~A.~Lineros}
\author[a]{A.~L.~Maroto}

\address[a]{Departamento de F\'{\i}sica Te\'orica I, Universidad Complutense de Madrid, E-28040 Madrid, Spain}
\address[b]{Instituto de F\'{\i}sica Corpuscular (CSIC-Universitat de Val\`{e}ncia), Apdo. 22085, E-46071 Valencia, Spain}

\begin{abstract}
We present a study of the Galactic Center region as a possible source of both secondary gamma-ray and neutrino fluxes from annihilating dark matter. We have studied the gamma-ray flux observed by the High Energy Stereoscopic System (HESS) from the J1745-290 Galactic Center source. The data are well fitted as annihilating dark matter in combination with an astrophysical background. 
The analysis was performed by means of simulated gamma spectra produced by Monte Carlo event generators packages. We analyze the differences in the spectra obtained by the various Monte Carlo codes  developed so far in particle physics. 
 We show that, within some uncertainty, the HESS data can be fitted as a signal from a heavy dark matter density distribution peaked at the Galactic Center, with a power-law for the background with a spectral index which is compatible with the Fermi-Large Area Telescope (LAT) data from the same region. If this kind of dark matter distribution generates the gamma-ray flux observed by HESS, we also expect to observe a neutrino flux. We show prospective results for the observation of secondary neutrinos with the Astronomy with a Neutrino Telescope and Abyss environmental RESearch project (ANTARES), Ice Cube Neutrino Observatory (Ice Cube) and the Cubic Kilometer Neutrino Telescope (KM3NeT). Prospects solely depend on the device resolution angle 
when its effective area and the minimum energy threshold are fixed. 
\end{abstract}


\end{frontmatter}



\section*{Introduction}
\label{1}

Astrophysical evidences for
Dark Matter (DM) exist from galactic to cosmological scales, but the interactions with ordinary matter have not
been probed beyond gravitational effects. In this sense, both direct and indirect DM searches
are fundamental to explore particle models of DM. If DM annihilate or decay in Standard Model (SM) particles, we may indirectly detect the secondary products of such processes in astrophysical sources where the DM density is dominant. In this context, the observation of secondary particles is highly affected by astrophysical uncertainties, such as the DM densities and distribution in the Galaxy and the astrophysical backgrounds. 
In particular, the Galactic Center (GC) represents an interesting source due to its closeness to the Earth, but also a complex region because of the large amount of sources present. In this work, we review the
analysis  of the data collected by the
HESS collaboration during the years 2004, 2005, and 2006 associated to the HESS J1745-290 GC gamma-ray source 
as a combination of a DM signal with an astrophysical power-law background. The best fits are obtained for the $u\bar u$ and $d\bar d$ quark channels and for the $W^+W^-$ and $ZZ$ gauge bosons with large astrophysical factors $\approx 10^3$  \cite{Cembranos:2012nj, HESS}. Such a parameter is affected not only by the astrophysical uncertainty, but also by the error introduced by the use of differential fluxes simulated by means of Monte Carlo event generator software. The exact estimation of the last effect depends on several factors, such as the annihilation channel, the energy of the process and the energy range of interest \cite{MC}.  In this contribution we focus on the $W^+W^-$ annihilation channel.  In addiction to the gamma rays study, we present some prediction on the prospective neutrino flux that may be originated by the same source. \\
This work is organized as follows. 
In the first section we revisit the equations able to describe both the gamma ray and neutrino fluxes from Galactic sources. 
The second section focuses on gamma rays phenomenology. There we show the fit of the HESS data for the $W^+W^-$ annihilation channel. Although the analysis is model independent, such annihilation channel possesses some interest for heavy dark matter models \cite{WIMPs}, such as branons among others \cite{branons}. In order to give an estimation of the error introduced by the Monte Carlo simulations, we analyze the case of photon spectra generated by both \pythia and \herwig packages, both in Fortran and C++. In particular we show results for  $2$ TeV center-of-mass events in the $W^+W^-$ channel (see \cite{MC} for more cases). In section 3, we consider the expected neutrino signal from the annihilation of  the heavy  DM required to produce the HESS gamma-ray signal. \\

\section{Astrophysical flux}
\label{2}
In general, both gamma ray and neutrino flux for one particular annihilation channel can be described by the equation for uncharged particles that travel without deviation due to galactic magnetic fields:    
\begin{equation}
\left(\frac{{\rm d}\Phi^{}_{}}{{\rm d}E_{}}\right)_j^i\,=\,\frac{\langle\sigma_i v \rangle}{8\pi M_{{\rm DM}_i}^2}\left( \frac{{\rm d}N}{{\rm d}E}\right)_j^i\times \langle J\rangle^i_{\Delta\Omega_j}\, \, {\rm GeV}^{-1}{\rm cm}^{-2}{\rm s}^{-1}{\rm sr}^{-1}\,,
\label{nuflux}
\end{equation}
where $j=\gamma,\nu_k$ is the secondary uncharged particle. When $j=\nu_k$, $k=\mu,\tau,e$ is the neutrino flavor. The DM annihilation channel is fixed by the $i=i$-th SM-particle. Because we performed single channel model independent fits, the astrophysical factor depends on the annihilation channel. Here, we present the results for the $i=W^{\pm}$ boson channel.
The differential number of particles ${\rm d}N/{\rm d}E$ is simulated by means of the Monte Carlo event generator software, as discussed in section $2.1$. Unlike  gamma rays,  the composition of the neutrino 
flux produced at the source can differ from that detected on the  Earth because of the 
 combination of different flavors produced by oscillations  \cite{Neutrinos}.\\


\section{Gamma-ray flux}
\label{2}

\begin{figure}[t]
\begin{center}
{\includegraphics{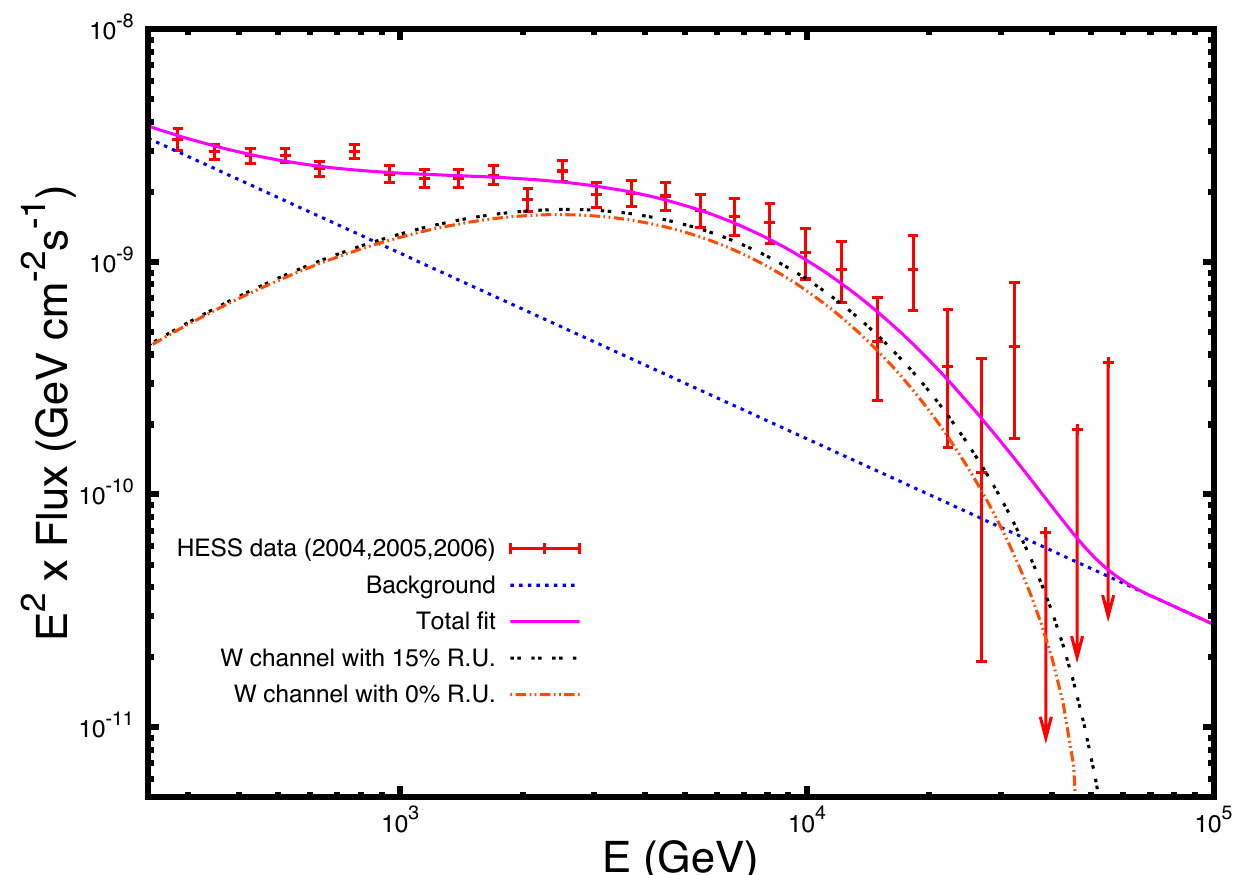}}
\caption {\footnotesize{
Best fit to the HESS J1745-290 collection of data \cite{HESS} in the case that the DM
contribution came entirely from annihilation into $W^+W^-$. The full line shows the total fitting function of $\chi^2=0.84$. The dotted line shows the simple power-law background component with spectral index $\Gamma=2.80\pm0.15$ and $B=5.18\pm2.23\times 10^{-4}{\rm GeV}^{-1/2}{\rm cm}^{-1}s$, while the dotted-dashed line shows the contribution of annihilating DM into $W^+W^-$ pairs without taking into account the telescope resolution uncertainty (R. U.) in energy. The two-dotted line takes into account such resolution in energy, that is a $15\%$. The fitted DM mass is $48.8$ TeV and the normalization parameter $A=4.98\pm0.40\times 10^{-7} \text{cm}^{-1}{\rm s}^{-1/2}$.
}}
\label{fig:fig1}
\end{center}
\end{figure}

As introduced before, the gamma rays signal observed by HESS between $200$ GeV and $10$ TeV from the GC direction may be a combination of a DM signal with a simple power-law background.
The total fitting function for the observed differential gamma ray flux is:

\begin{equation}
\frac{{\rm d}\Phi_{\gamma-Tot}}{{\rm d}E}=\frac{{\rm d}\Phi_{\gamma-Bg}}{{\rm d}E}+\frac{\Phi_{{\rm d}\gamma-DM}}{dE}=B^2\cdot \left(\frac{E}{\text{GeV}}\right)^{-\Gamma}+ A_i^2 \cdot \frac{{\rm d}N^i_{\gamma}}{{\rm d}E}\,,
\label{gen}
\end{equation}
where
\begin{equation}
\label{A}
A_i^2=\frac{\langle \sigma_i v \rangle\, \Delta\Omega_\gamma^{{\rm HESS}}\, \langle J \rangle_{\Delta\Omega_\gamma^{{\rm HESS}}}}{8\pi M_{\rm DM}^2}
\end{equation}
needs to be fitted together with the DM particle mass $M_{\rm DM}$, the background amplitude $B$ and
spectral index $\Gamma$. By means of the fit of the parameters $A_i$, the astrophysical factor  
\begin{eqnarray}
{\langle J \rangle}^i_{\Delta\Omega}\,=\, \frac{1}{\Delta\Omega}\int_{\Delta\Omega}\text{d}\Omega\int_0^{l_{max}(\Psi)} \rho^2 [r(l)] \,{\rm d}l(\Psi)\,,
\label{J}
\end{eqnarray}
is also indirectly fitted. In the previous expression, $l$ holds for the distance from the Sun to any point in the halo. It is related with the radial distance $r$ from the GC as $r^2 = l^2 + D_\odot^2 -2D_\odot l \cos \Psi$, where $D_\odot \simeq 8.5$ kpc is the distance from the Sun
to the center of the Galaxy. The maximum distance from the Sun to the edge of the halo in the direction $\Psi$ is
$l_{max} = D_\odot \cos\Psi+ \sqrt{r^2-D_\odot^2 \sin \Psi}$. Moreover, the photon flux must be averaged over the solid angle of the detector. For the HESS telescope observing gamma rays in the TeV
energy scale, it is of order
$\Delta \Omega_\gamma^{\rm HESS} = 2 \pi ( 1 - \cos\theta) \simeq 10^{-5}$. The DM density distribution in the Galaxy halo  is usually modeled by the NFW profile \cite{Navarro:1996gj}:
\begin{equation}
\rho(r)\equiv\frac{N}{r(r-r_s)^2}\;,
\label{NFW}
\end{equation}
where $N$ is the overall normalization and $r_s$ the scale radius. This profile is in good agreement with N-body non-baryonic cold DM simulations of the GC.
In this case and accounting for just annihilating DM, the astrophysical factor is:
$\langle J^{\text{NFW}} \rangle\simeq 2.8 \cdot 10^{25}\; \text{GeV}^2 \text{cm}^{-5}$. We will use this value as standard reference in order to define the boost factor as  $b^i\equiv \langle J\rangle^i/\langle J^{\text{NFW}}\rangle$. The differential number of gamma photons ${\rm d}N^i_\gamma/{\rm d}E$ is simulated by \pythia 6.4 and the analytical fitting functions of such simulations are used to get the fit \cite{pythia6, Ce10}. In Fig. \ref{fig:fig1}  and Table \ref{tab:tab1} we report the results of the gamma rays fit of the $W^+W^-$ channel. The fit well reproduces the spectral form and the energy cut-off  for a DM mass of $\sim 50$ TeV and a boost factor of $\sim 10^3$. The background is also fitted and its amplitude and spectral index are in agreement with the FERMI-LAT data \cite{Fermi}. The uncertainty introduced by the choice of the simulator is discussed in the next section. 

\begin{table}
\begin{center}
 \begin{tabular}[h!]{cc}
  \hline
    \hline
   Channel &     $W^+W^-$     \\ \hline
    \hline
    $M$  &  $48.8\pm 4.3$ \\ 
    $A$  & $4.98\pm0.40$\\  
    $B$ & $5.18\pm2.23$\\  
    $\Gamma$ & $2.80\pm 0.15$\\
    $\chi^2/\,$d.o.f. & $0.84$\\
    $\Delta\chi^2$ & $2.6$\\
    $b$ & $1767 \pm 419$\\
    \hline
    \hline
\end{tabular}
\caption{\footnotesize{Fit parameters for the gamma-ray flux from the GC region HESS J1745-290 as DM annihilating into $W^+W^-$ channel combined with the diffuse background. The DM parameters $M$ and $A$ represent the DM mass and the amplitude of the signal, respectively. The power-law background is described by its amplitude $B$ and spectral index $\Gamma$. The normalized $\chi^2$/d.o.f. and its deviation from the best fit $d\bar d$ channel (see \cite{HESS}) is given. Finally, $b$ is the boost factor with respect to a NFW DM density profile.}}
\label{tab:tab1}
\end{center}
\end{table}

\subsection{Monte Carlo uncertainty on photons flux}

The differential number of photons simulated by Monte Carlo event generators software is the result of a particle shower schematization in three parts: the QCD Final-State Radiation, the hadronization model and the QED Final-State Radiation.
The four codes differ in each one of the aforementioned fundamental parts and the combination of these intrinsic differences to the final spectra is complicated (see ~\cite{MC, Seymour:2013ega} for details). In any case, it seems clear that the parton shower evolution variable affects the QCD Final-State Radiation ~\cite{Beringer:1900zz, Altarelli:1977zs, 81, 83, CMW, Py6}, while the hadronization model (String model in $\text{\pythia}\text s$ ~\cite{Py6, Py8} and Cluster model in $\text{\herwig}\text s$ ~\cite{Her, Her++}) produces unstable hadrons which eventually decay. The resultant final states of such process are mainly leptons leading the photon production that involves QED processes. Finally, the Bremsstrahlung component of the Final-State Radiation (FSR) represents the main difference between the codes, for photons spectra at least. In fact, the Bremsstrahlung of very high energy leptons is not implemented in \herwig C++, it is just partially implemented in \herwig Fortran, while it is implemented both in \pythia++ and \pythia Fortran. The electroweak (EW) $2\rightarrow2$ processes of the FSR, despite their different implementations for different codes, do not significantly affect the gamma-ray spectra.\\
\begin{figure}[tb]
\centering
\resizebox{\columnwidth}{!}
{\includegraphics[height=100pt]{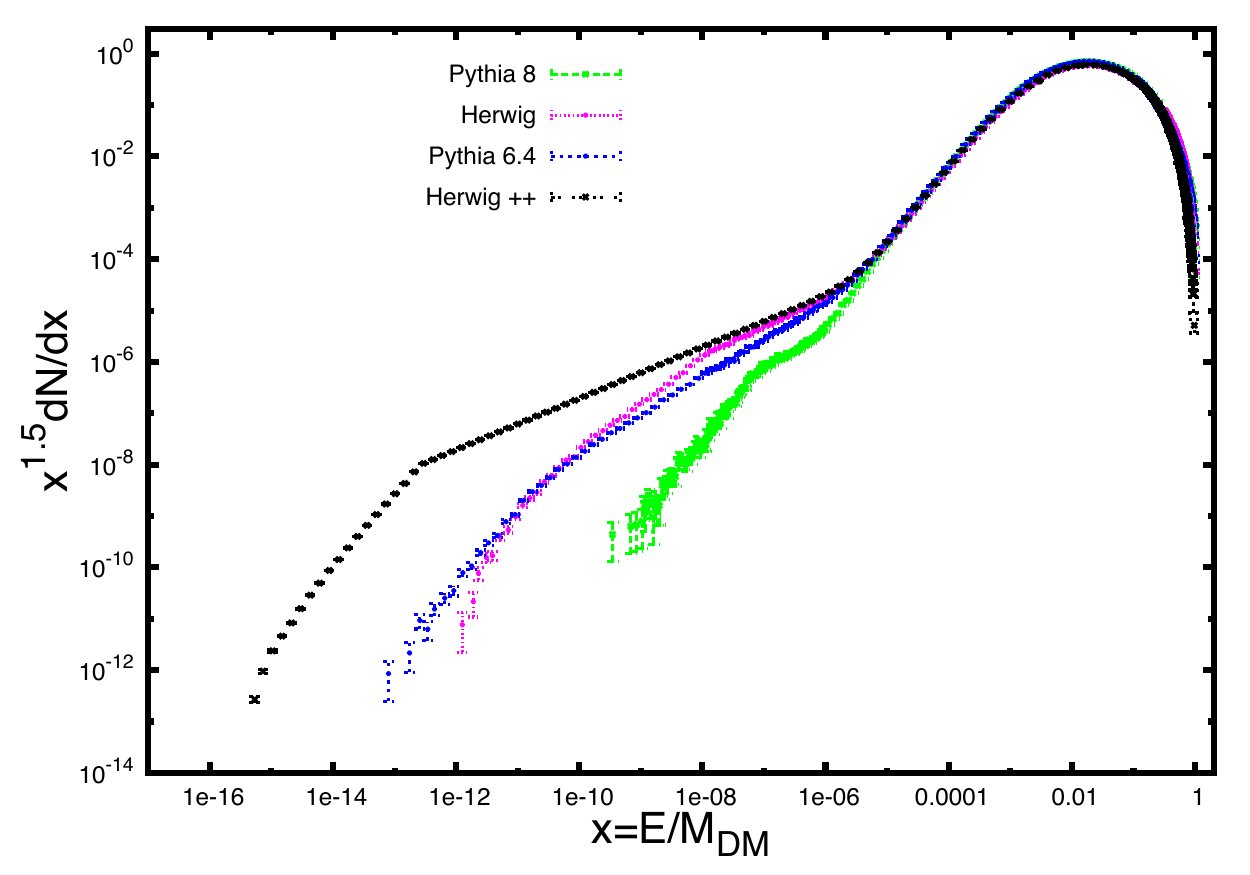}\includegraphics[height=100pt]{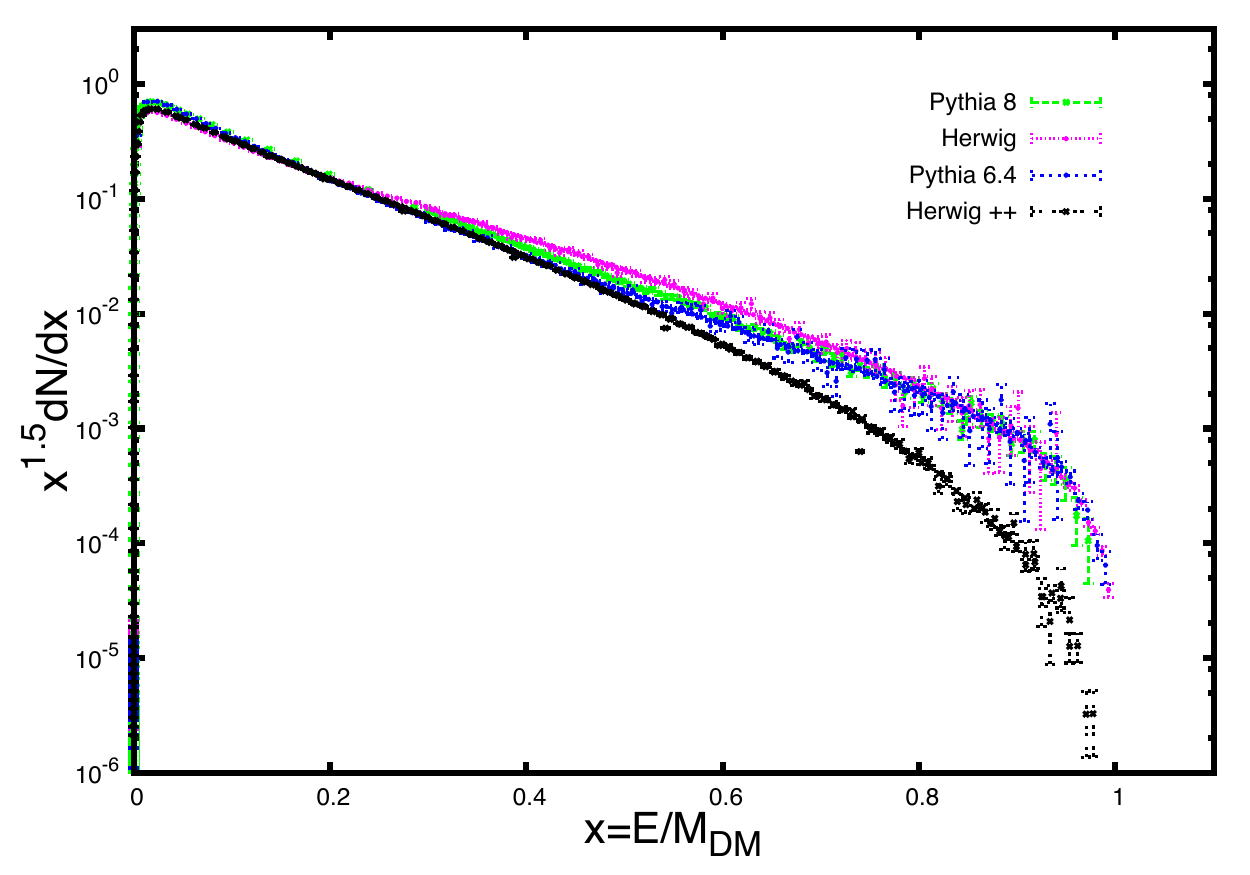}}
\caption{
 $W^+W^-$ annihilation channel with $M_{\rm DM}=1$ TeV. {\it (Left--panel)} Logarithmic scale: the simulations are consistent down to a value of $x\approx10^{-6}$.
{\it (Right--panel)} Linear scale: all the  simulations exhibit the same behavior, except for \herwigpp that separates from the other simulations for values above $x\simeq0.3$.}
\label{fig:w1000g}
\end{figure}

\label{3}
%
%

For annihilating DM, the photon spectra are better described in terms of the dimensionless variable:
	$x \equiv E_{\gamma}/E_{\rm CM} $
where $E_{\gamma}$ and $E_{\rm CM}$ correspond to the energy of the photon and center of mass (CM), respectively. This variable lies in the range between 0 to 1. 
Because the standard Monte Carlo adjust uses data at $E_{\rm CM}=100$ GeV of center of mass energy from colliders such as LEP and LHC, large differences in the spectra are usually present at very low or very high values of $x$. For this reason, we present the spectra in both linear and logarithmic scales for $x$, in order to show more clearly the behavior in the first and second case, respectively. Here, we focused on DM particle mass of $1$~TeV (see \cite{MC} for the same analysis with $100\,\,\text{GeV}$ DM annihilating into $W^+W^-$, $b\bar{b}$, $\tau^{+}\tau^{-}$ or $500$ GeV DM mass in the case of $t\bar t$ and further details on $1$ TeV DM mass for more channels). The photon spectra are independent of the initial beams (details on the generation of the spectra can be found in \cite{MC}) and solely depend on the energy of the event, i.e. $E_{\rm CM} = 2 M_{\rm DM}$ for annihilating DM. 
 In Fig. \ref{fig:w1000g} we show that the simulated gamma-ray spectra for DM particles annihilating in $W^+W^-$ are very similar for $ x \gtrsim10^{-5}$ for a DM mass of $1$ TeV. The lower fluxes generated by \herwigpp at high
energies (linear scale) are probably due to the lack in the implementation of the Bremsstrahlung radiation from high energy leptons. \\

\begin{figure}[tb]
\centering
\includegraphics{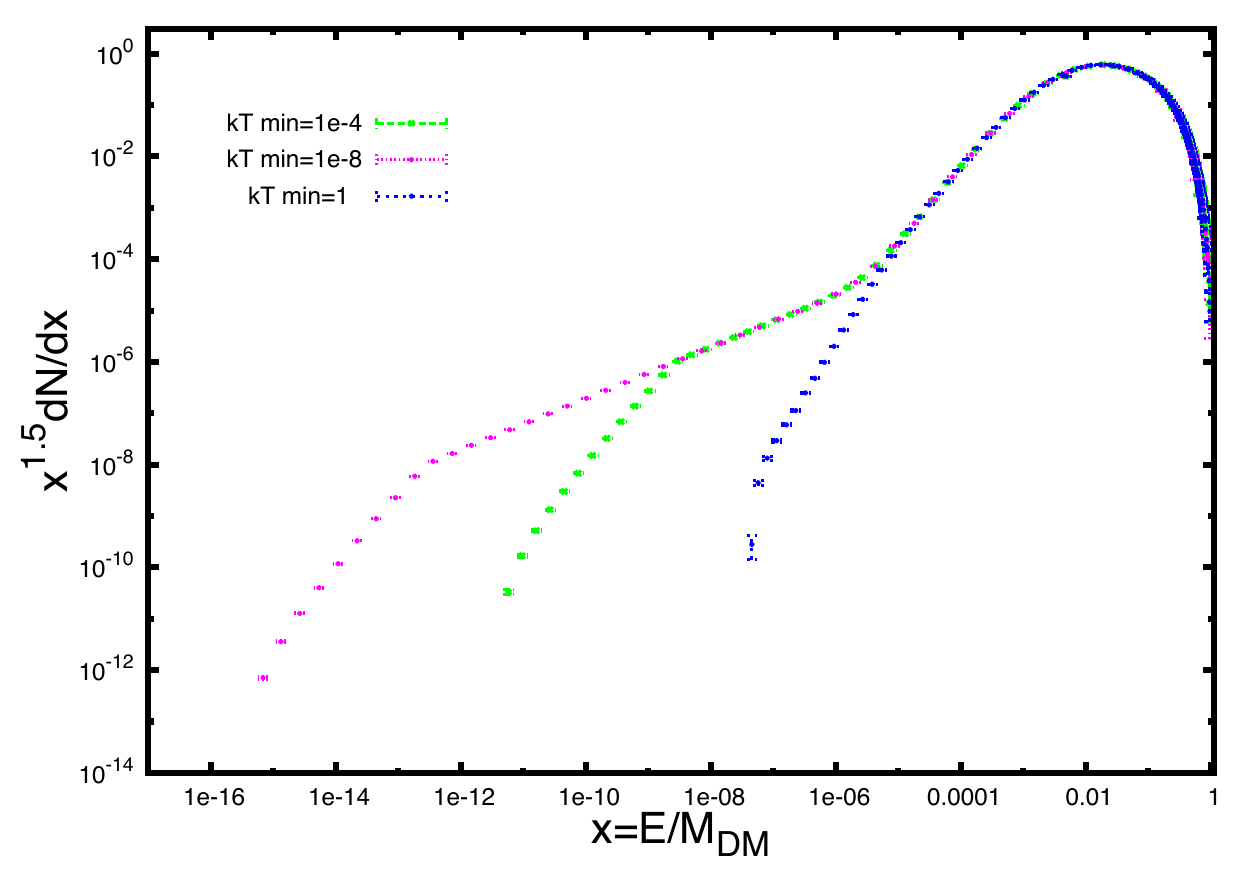}
\caption{
The difference at low energy in C++ codes can be explained by the parameters that cut-off the lower energy photons. High energies can be proved not to be affected by this fact. $W^+W^-$ channel with \herwig++ at $M_{\rm DM}=1$ TeV in logarithmic scale. The three cuts-off correspond to \texttt{QEDRadiationHandler} of $k_T=10^{-8}\,, 10^{-4}\,, 1$.}
\label{fig:w1tevqed}
\end{figure}

Although the low energy spectra are less important in the context of indirect searches due to the dominance of the astrophysical background, let us underline that the low energy cut-off strongly depends on the set parameters of each software. In \pythia 8 the cut-off at low energy exactly corresponds to the minimum value allowed for photons and set by the \texttt{pTminChgL} parameter, that is 0.005 by default.
In \herwigpp, \texttt{QEDRadiationHandler} is set off by default, so that the cut-off appears at higher energy with respect to the other Monte Carlo generators.
In the opposite case, when \texttt{QEDRadiationHandler} is enable, the spectrum at low energy changes drastically. In this case, the relevant parameter \texttt{IFDipole:MinimumEnergyRest} can be varied in values: small values of such parameter enlarge the production of photons at low energies. Three different low energy cutoffs for the $W^-W^-$ channel are shown in Fig. \ref{fig:w1tevqed}.\\

\begin{figure}[tb]
\centering
\resizebox{\columnwidth}{!}
{\includegraphics[height=100pt] {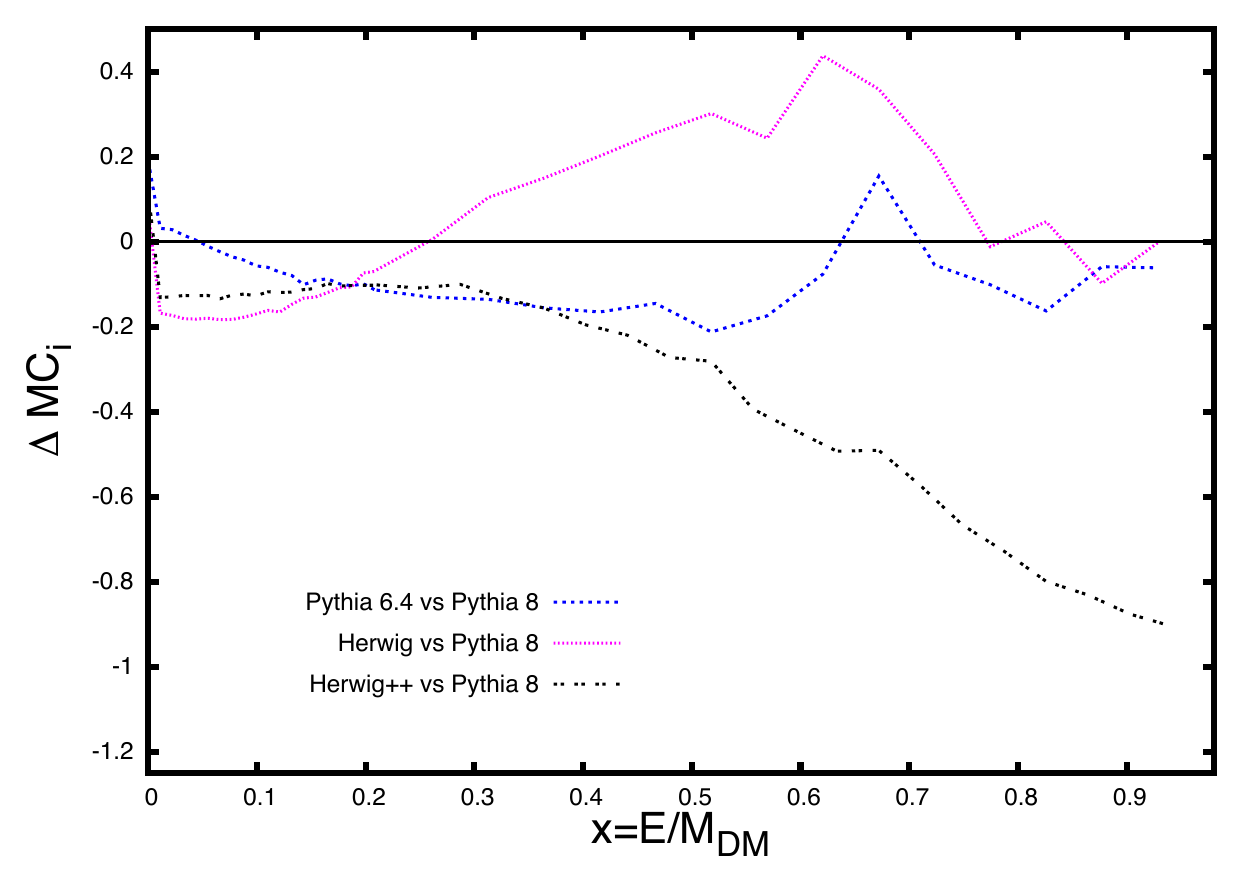}\includegraphics[height=100pt]{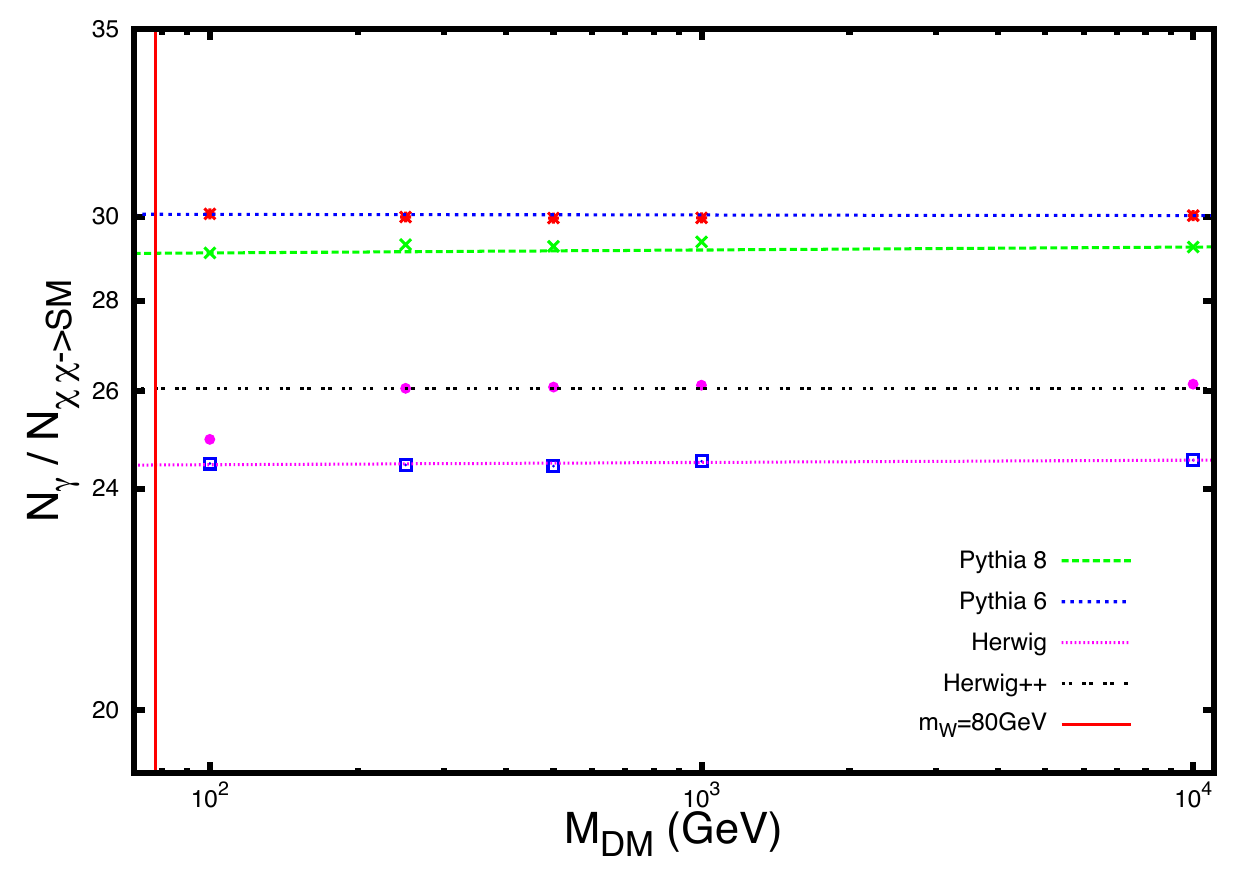}}
   \label{ER_w}
   
\caption{
Left Panel: Relative deviations versus $x$ at $M_{\rm DM}=1$ TeV.
The full horizontal line at zero represents \pythia 8.
The dashed blue line holds for \pythia 6.4 vs. \pythia 8,
the dotted one is \herwig Fortran vs. \pythia 8 and
the two-dotted one is \herwigpp vs. \pythia 8. Right Panel: Multiplicity with lower photon energy cut-off of $x_C = 10^{-5}$.  \pythia 6.4 provides the upper limit to the number of generated photons by the $W^+W^-$, while \herwig Fortran provides the lower limit with $23\%$ difference between them.
}
\label{ErrRel}
\end{figure}

More interesting are the spectra at high energies.
%
We present the Monte Carlo relative deviation ($\Delta {\rm MC}_i$) with respect to \pythia 8 in Fig. \ref{ErrRel}, defined as
\begin{equation}
\Delta {\rm MC}_s =
\;\frac{{\rm MC}_s\, - {\text{ \pythia}} \, 8}
{{\text{ \pythia}} \, 8},
\label{RelErr}
\end{equation}
where $MC_s$ stands for \pythia 6.4, \herwig and \herwigpp.
For a DM mass of $1$ TeV and $W$ boson annihilation channel, the relative deviations for $ x\gtrsim0.01$  are always less than $20\%$ up to $x=0.2$. At $x\gtrsim 0.2 $ the absence of Bremsstrahlung radiation generated by high energy leptons in \herwig++ leads to a smaller number of high-energy photons when compared to the other softwares. 

Also the multiplicity, that is the total number of photons produced by each event, affects the constraints.
Apart from the specific characteristics of the detector, the flux of photons depends upon the DM density distribution and the distance of the sources. Thus, two simulations should give different number of photons for the same number of events and this will affect the parameters $\langle J\rangle$ and $b$.
In general, the multiplicity depends on both the Monte Carlo event generator, the energy of the event and the annihilation channel. For the $W$ boson channel, it does not depend on the mass above $300$ GeV, as we can see in Fig. \ref{ErrRel}. In this study, the energy cut-off increases with the DM mass, because we set a lower photon energy cut-off around  $x_C=10^{-5}$. This kind of DM mass depending cut-off allows to reject photons of lower energies, where the simulations present important differences, and the contribution to gamma rays is less important. This cut-off is also compatible with typical gamma-ray detectors energy thresholds, in fact detector energy thresholds are typically around $1-10$ GeV depending on the particular experimental device \cite{branonsgamma}. In any case, our results do not seem to depend on the particular choice of this cut-off. 
The multiplicity behavior is well approximated by the following power law relation with the DM mass:
\begin{equation}
\frac{N_{\gamma}}{N_{\chi\chi \rightarrow SM}}\simeq a\cdot \left(\frac{M}{1\,\text{GeV}}\right)^b\;.
\label{pl}
\end{equation}
The $a$ and $b$ coefficients are given in \cite{MC}. They depend on both the Monte Carlo simulator and the annihilation channel. When the SM particle is fixed, cosmological constraints obtained by means of  the total number of generated gamma photons might depend on the Monte Carlo simulation.\\
In the case of the $W^+W^-$ and whenever the DM annihilation cross section is fixed, \pythia 6.4 provides lower limit values for astrophysical factor/boost factor in gamma rays, depending on the kind of fit. On the other hand, \herwig Fortran gives the upper limit for similar analysis (see Fig. \ref{ErrRel} and \cite{MC} for details). In any case, the difference between \pythia 6.4 and \pythia 8, the latter being 
the most complete Monte Carlo software for gamma rays, is less that a $4\%$.

\begin{figure}[t]
\begin{center}
{\includegraphics{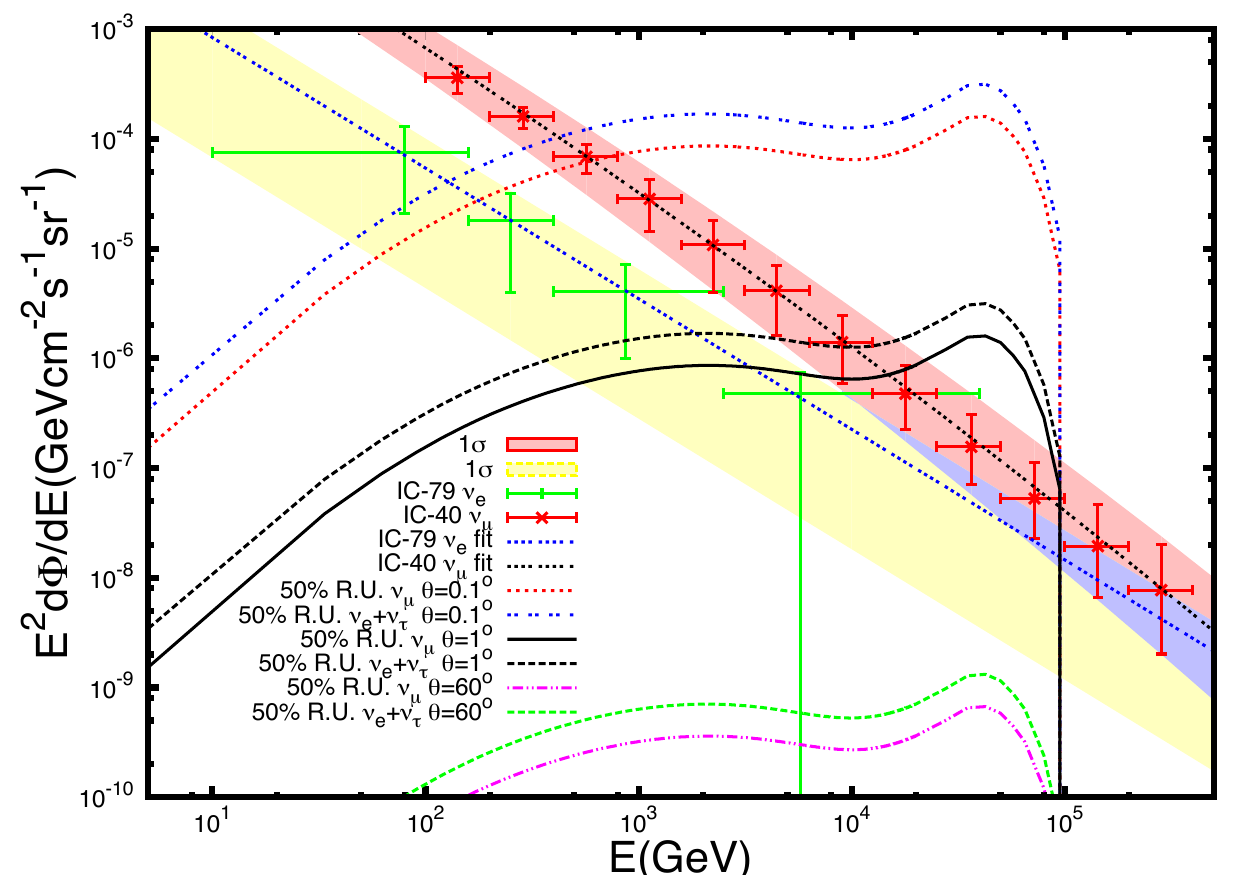}}
\caption {\footnotesize{Expected neutrino fluxes from DM annihilating into $W^+W^-$ channel for three different resolution angles, namely $\theta=60^\circ, 1^\circ, 0.1^\circ$. For each couple of lines (each for one value of $\theta$), the upper dot line is the $\nu_e$ expected flux on Earth. Due to the fact that the detector is able to distinguish between both  $\nu_e$-$\nu_\tau$ and $\nu_i$- $\bar\nu_i$, with the $\nu_e$ flux we have performed the sum of all them. The lower full line of each pair is the same for $\nu_\mu$. The muonic and electronic background are given by IceCube 40-string configuration and IceCube in the 79-string configuration, respectively. }}
\label{Wnu}
\end{center}
\end{figure}

\section{Neutrino flux}
\label{3}

\begin{figure}[h!tb]
\begin{center}
\resizebox{9.0cm}{9.5cm}
{\includegraphics{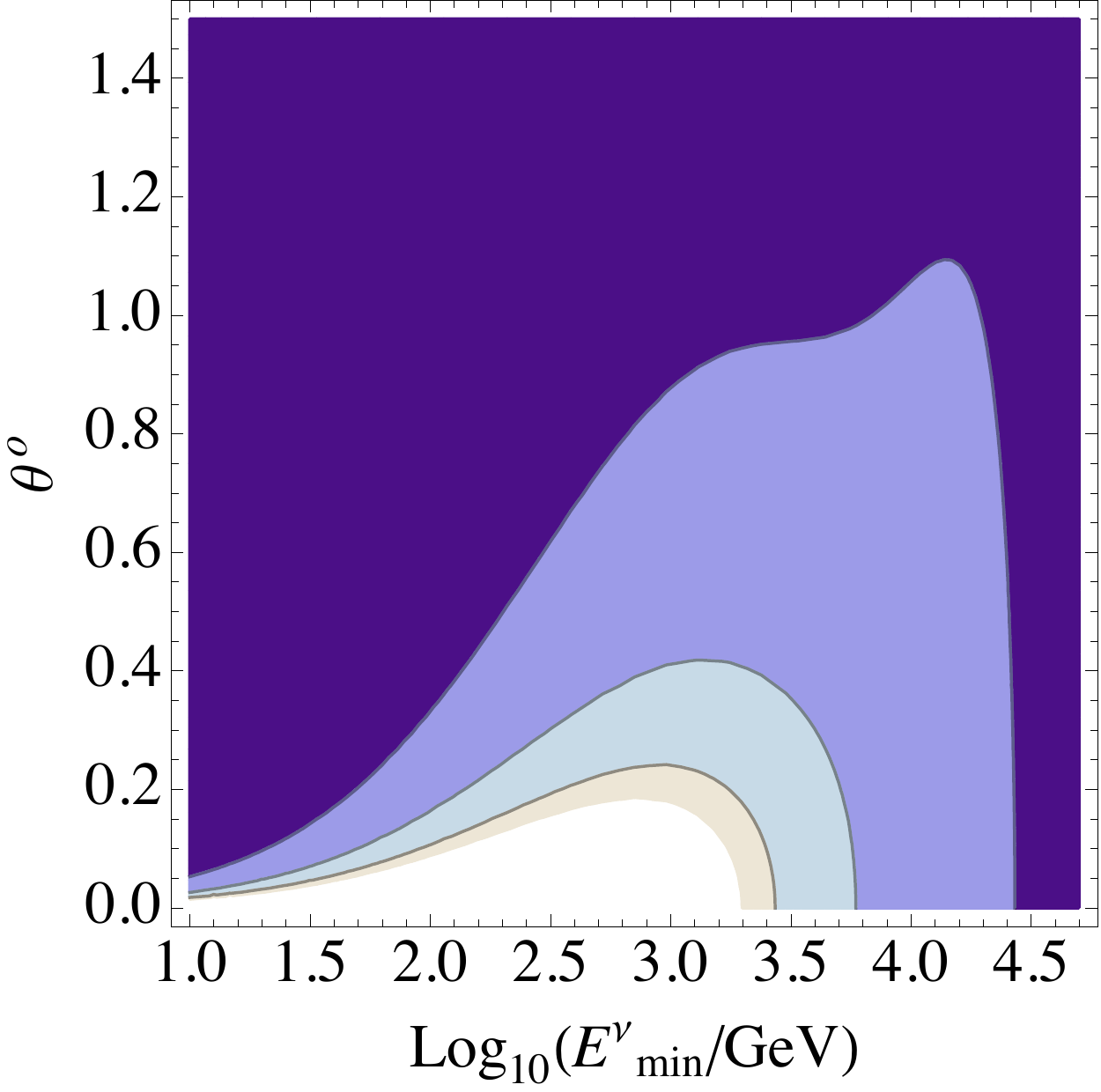}}
\caption {\footnotesize{The Figure shows the $1\sigma$ (dark), $2\sigma$, $3\sigma$, $5\sigma$ (white) confidence levels contours
in the case of DM annihilating into the $W^+W^-$ channel. The factor $A_{\text{eff}}\times t_{\text{exp}}=100\, \text{m}^2\,\text{yrs}$. The possibility of detecting the neutrino flux signal
above the atmospheric background depends on the energy cut $E_{\text{min}}^\nu$ and the resolution angle.}}
\label{Aeff}
\end{center}
\end{figure}

As for gamma rays, the differential neutrino flux from annihilating DM in the Galaxy is described by the equation (\ref{nuflux}).
As a difference with the gamma ray case, the neutrino flux on the Earth is not the same as at the source, due to neutrinos oscillations and observational limits. In fact, neutrino telescopes are able to discriminate between $\nu_\mu$ and $\nu_e$ or  $\nu_\tau$, but they cannot identify either $\nu_e$ from $\nu_\tau$ or neutrinos from anti-neutrinos. So, the interpolation functions of the differential neutrinos spectra simulated by \pythia 8 \cite{Py8, Cirelli} need to be slightly modified.  In Fig. \ref{Wnu} we show the expected neutrino flux on Earth when neutrino oscillations and observational limits are taken into account. Because all the parameters of the model are fitted by the observation in gamma rays, observability of the neutrinos flux depends only on a combination of the resolution angle and effective area of the telescope with the minimum energy threshold and the exposition time. Both the angle and the area depend on the neutrino flavor and the position of the source in the sky with respect to the detector \cite{ANTARES, IC, km3net}. In the case of the GC, it should be possible in principle to get better resolution angle with ANTARES \cite{ANTARES} than with IceCube \cite{IC}. In the first case the Earth is used as veto and the background is given by atmospheric neutrinos. On the other hand, when the background is given as most by atmospheric muons, as in the case of IceCube, the effective area of the detector is affected. 
The IceCube collaboration reports the $\nu_\mu$ and $\nu_e$ atmospheric neutrino fluxes \cite{nue, numu}.  As is shown in Fig. \ref{Wnu}, no flux is expected with resolution lower that $\theta\approx1^\circ$ from high density $48.8$ TeV DM annihilating into $W^+W^-$ channel in the GC. Thus we need better resolution angles in order to be able to get some signature above the background. In fact, the statistical significance of the signal above the background is given by
\begin{equation}
\chi_{\nu_k}=\frac{\Phi_{\nu_k}\sqrt{A_{\text{eff}}\,t_{\text{exp}}\,\Delta\Omega}}{\sqrt{\Phi_{\nu_k}+\Phi^{\text{Atm}}_{\nu_k}}} = 5\, (3,\, 2)\,.
\label{Chi1}
\end{equation}
$\chi_{\nu_k}$ depends also on the minimum energy threshold, exposition time and effective area.
In Fig. \ref{Aeff} we show the statistical analysis for $\nu_\mu$ track events for a generic neutrino telescope with $A_{\rm eff}\times t_{\rm exp}=100\,\text{m}^2\,yr$. So, with $A_{\rm eff}=20\,\text{m}^2$, $t_{\rm exp}=5\, {\rm yrs.}$ and $E_{min}=1$ TeV, an angle $\theta\leq0.4^\circ$ is requested to get a signal measurement with a confidence level better than $2\sigma$. Similar analysis was developed fixing the resolution angle and searching for the $A_{\rm eff}\times t_{\rm exp}$ parameter with respect to the minimum energy threshold \cite{Neutrinos}.


\section{Conclusions and Prospects}

We have analyzed the gamma-ray and neutrino flux that should be generated by a very peaked DM distribution in the GC and presented partial results for the $W^+W^-$ annihilation channel. The study is based on the fit of the HESS data in gamma rays allowing to constrain the DM mass and the astrophysical factor. Among other channels, the collection of data of the Cerenkov detector for the J1745-290 source is well fitted as $48.8$ TeV DM annihilating into $W^+W^-$ boson particles. The signal is  superimposed on a gamma-ray background compatible with the Fermi-LAT observation. We have also analyzed  the uncertainty that may be introduced by the simulation of gamma-ray flux with different Monte Carlo particle physics codes. The relative deviation between different codes turns out to be less than $20\%$ for the boson channel at the energy range of interest, whereas the number of photons produced for each event introduces an error less than $4\%$. These uncertainties may affect the $10^3$ enhancement of the astrophysical factor necessary to fit the data. The astrophysical factor may also be affected by the astrophysical uncertainty due to the choice of the DM density profile. In any case, its value is compatible with the baryonic enhancement in Monte Carlo cosmological N-body simulation \cite{Blumenthal, Prada:2004pi},  although opposed opinions about this scenario remain \cite{Romano}. 
For the DM particle able to fit the gamma-rays data, we have also presented the prospects for  the detection of the neutrinos flux to be generated
by that particle. It depends both on the resolution angle and effective area of the neutrino telescope, in addition to the minimum energy threshold and the observation time. We sketched a partial study of the combined resolution angle and the energy threshold needed to detect a neutrino signal within some confidence level, when the effective area and the exposition time are fixed. A resolution angle of $0.4^\circ$ is requested to get $2\sigma$ signal above the background for neutrino telescope with effective area compatible with ANTARES or IceCube, but more exposition time than we have available with the actual collection of data \cite{nue, numu} is required. 
Therefore, at the present stage we are not able to either accept nor reject the DM origin for gamma rays data with the present generation of neutrino detectors. An improvement in the angular resolution of ANTARES or IceCube when looking at the GC may be fundamental in order to clarify DM hypotheses. Moreover,  the observation of this region with the next KM3NeT neutrino detector \cite{km3net}, with an effective area of $1\,\text{km}^2$ and improved resolution angle, will be also of great interest. 
Finally, the observation of antimatter flux and matter-antimatter ratio as proton, antiproton and positron signal, may be useful to set additional constraints on the DM origin. 



\vspace{5 cm}

\section*{Acknowledgments}

This work was supported by UCM FPI grants G/640/400/8000 (2011 Program), the Spanish MINECO 
projects numbers FIS2011-23000, FPA2011-27853-C02-01and  MULTIDARK CSD2009-00064 (Consolider-Ingenio 2010 Programme).
A.d.l.C.D. also acknowledges the hospitality of IFAE-UAB in the final stages of preparation of this manuscript.








\end{document}